\documentclass[twocolumn,showpacs,preprintnumbers,amsmath,amssymb,aps,prl,
]{revtex4}

\def\aap{{A\&A}}

\def\pra{{Phys.~Rev.~A}}
\def\prb{{Phys.~Rev.~B}}

\def\pre{{Phys.~Rev.~E}}

\def\jcp{{J.~Chem.~Phys.}}

\usepackage{upgreek} 

\usepackage{epsfig, amsmath,amsfonts, amssymb,graphicx,amsthm,color}

\usepackage{mathtools}
\usepackage{dcolumn}
\usepackage{bm}
\usepackage{rotating}
\usepackage{amssymb}
\usepackage[latin1]{inputenc}
\usepackage{graphicx}

\usepackage{xr}
\usepackage{epstopdf}
\newcolumntype{M}[1]{>{\raggedright}m{#1}}

\begin{document}

\title{Phase space density limitation in laser cooling without spontaneous emission}
\author{Thierry Chaneli\`{e}re$^*$, Daniel Comparat$^*$ and Hans Lignier$^*$}
\affiliation{Laboratoire Aim\'{e} Cotton, CNRS, Univ. Paris-Sud, ENS Paris Saclay, Universit\'e Paris-Saclay, B\^{a}t. 505, 91405 Orsay, France \\
\textrm{$^*$These authors contributed equally to this work}}

\date{\today}


\begin{abstract}
We study the possibility to enhance the phase space density of non-interacting particles submitted to a classical laser field without spontaneous emission.  
 We clearly state that, when no spontaneous emission is present, a quantum description of the particle motion is more reliable than semi-classical  description which can lead to large errors especially if no care is taken to smooth structures smaller than the Heisenberg uncertainty principle. 
Whatever the definition of position-momentum phase space density, its gain is severely bounded especially when started from a thermal sample. More precisely, the maximum of the  position-momentum phase space density, can  only increase by a  factor $M$ for $M$-level particles. This bound comes from a transfer between the external and internal degrees of freedom. Therefore, it is impossible to increase the position-momentum phase space density in the same internal state. 
\end{abstract}

\maketitle

It is usually believed that the phase space density (PSD) of non-interacting particles cannot be increased by using only pure Hamiltonian evolution and any PSD increase would require a dissipative mechanism  \cite{1992PhRvA..46.4051K,1997JChPh.106.1435B}. In the context of laser cooling, this dissipation is usually ensured by spontaneous emission.
Nevertheless, in recent years, several papers suggested that some experimental observations could support the possibility of an optical cooling without spontaneous emission \cite{2015PhRvL.114d3002C,Bartolotta2018,2018arXiv180504452G}. These counter intuitive results were also supported by theoretical arguments and semi-classical simulations using classical laser fields \cite{PhysRevA.54.R1773,2015JOSAB..32B..75C,PhysRevA.85.033422}. The perspective of cooling different species including molecules has actively stimulated the discussions  \cite{2008PhRvA..77f1401M,RevModPhys.89.041001,2018NJPh...20b3021N}.

In this letter, we specifically address the issue of increasing the PSD for non-interacting particles  submitted to classical laser fields (i.e. equivalent to quantum fields in coherent mode  \cite{mollow1975pure,Cohen-Tannoudji1997,dalibardcoolegeFrance2015}) and deprived of spontaneous emission. We first determine the evolution of a position-momentum distribution (PMD) of such particles (often called atoms hereafter eventhough molecules are also concerned) in a phase space region. In particular, we show that a quantum treatment of the external degrees of freedom is more reliable than a classical treatment that may lead to erroneous predictions. A quantum description of position and momentum requires to revisit the definition of the classical PMD, to define quantum analogs and to discuss their characterizations.
Because the term "cooling" is ambiguous and has often led to misinterpretations and controversies, we perform our analysis by considering both the PMD and several definitions of a single quantity (rather than a distribution) called PSD in a generic way. Somehow, the most straightforward definition of PSD is the maximum of PMD. Other definitions, such as those derived from different entropies, are used to account for the populations and correlations of the internal and external degrees of freedom. With these careful definitions, we establish that PSD can marginally increase in the case of an initial thermal distribution. Yet the gain is shown to be bounded by the number $M$ of internal levels.

First of all, it is important to recall that the evolution of non-interacting particles can be derived from a single particle statistics. In this framework, we do not study single realizations of many-particle evolution that may cause PMD modification driven by ergodicity, Zermelo-Poincaré recurrence or Fluctuation theorems \cite{2002AdPhy..51.1529E} as through coarse grained PMD \cite{2005PhRvA..72a3406P,2005A&A...430..771C} or by phase-space  volume surrounding particles   (such as ellipsoid emittance growth in  beams)   \cite{2003PhRvS...6c4202F}.  
Therefore, we assume the ensemble evolution as entirely derived from the one-particle density matrix $\hat \rho$ in the quantum case and, in the classical case, from  the (statistical averaged single particle) classical PMD $\rho (\bm r,\bm v,t)$.

The most general evolution of the classical PMD undergoing a (non-random) external force $\bm F (\bm r,\bm v,t)$ is given by the continuity equation:
\begin{equation}
\frac{D  \rho}{D t} =	\frac{\partial  \rho}{\partial t} + \left(\bm v \cdot \frac{\partial }{\partial \bm r} \right)  \rho + \frac{\bm F}{m} \cdot   \frac{\partial  \rho}{\partial \bm v} = - \rho \frac{\partial    }{\partial \bm v}.\frac{\bm F}{m}
\label{continuity_eq}
\end{equation}	
where $\frac{D \rho}{D t}$ is the material derivative. This clearly shows that a velocity-dependent force is necessary to change the PMD $\rho$. The  Doppler cooling scheme, using for example the classical Lorentz oscillator model, is a textbook example of velocity-dependent force.
However, in Hamiltonian mechanics, according to the Vlasov-Liouville's theorem $\frac{D  \rho}{D t} = 0$ for non-interacting particles, $\rho$ is constant. This is consistent with the continuity equation because friction forces cannot be included in our closed system with external fields \footnote{For example, an electric charge submitted to the Lorentz Force ${\bm F}= q({\bm E} + {\bm v} \times {\bm B})$ verifies $\frac{\partial}{\partial \bm v}. \bm F =0$}. Since quantum mechanics is also based on a Hamiltonian description, one may wonder how the maximum of a PMD could be increased. A major difference actually comes from the treatment of the internal degrees of freedom that cannot be rigorous in classical physics. Regarding the electromagnetic interactions, the time evolution of the internal degrees of freedom is generally calculated by the quantum master equation acting on the density matrix because it may also include non-unitary evolutions due to spontaneous emission. The semi-classical evolution of the external degrees of freedom is then usually obtained by Ehrenfest's theorem. This framework provides satisfying predictions for Doppler cooling where the change of semi-classical PMD maximum is essentially attributed to spontaneous emission.  However even without spontaneous emission, several semi-classical studies suggest that the maximum of a PMD can be modified ($\pi$-pulse, rapid adiabatic passage (RAP), Stimulated RAP, bichromatic fields \cite{PhysRevA.54.R1773,PhysRevA.85.033422,2015PhRvL.114d3002C,RevModPhys.89.041001,2018arXiv180504452G}). Their common idea is that a coherent force, resulting from absorption and stimulated emissions, depends on the particle velocity via the Doppler effect. So a large increase of the PMD maximum seems possible from the continuity equation (\ref{continuity_eq}).
 In the following, we will show that the concept of semi-classical force is only partly correct and that the Ehrenfest's theorem can lead to an important overestimation of the cooling efficiency. We will show that a proper quantum mechanical treatment exhibits a limited gain in the PMD, its maximum being the number $M$ of internal levels.

 \begin{figure}
 	\includegraphics[width=\columnwidth]{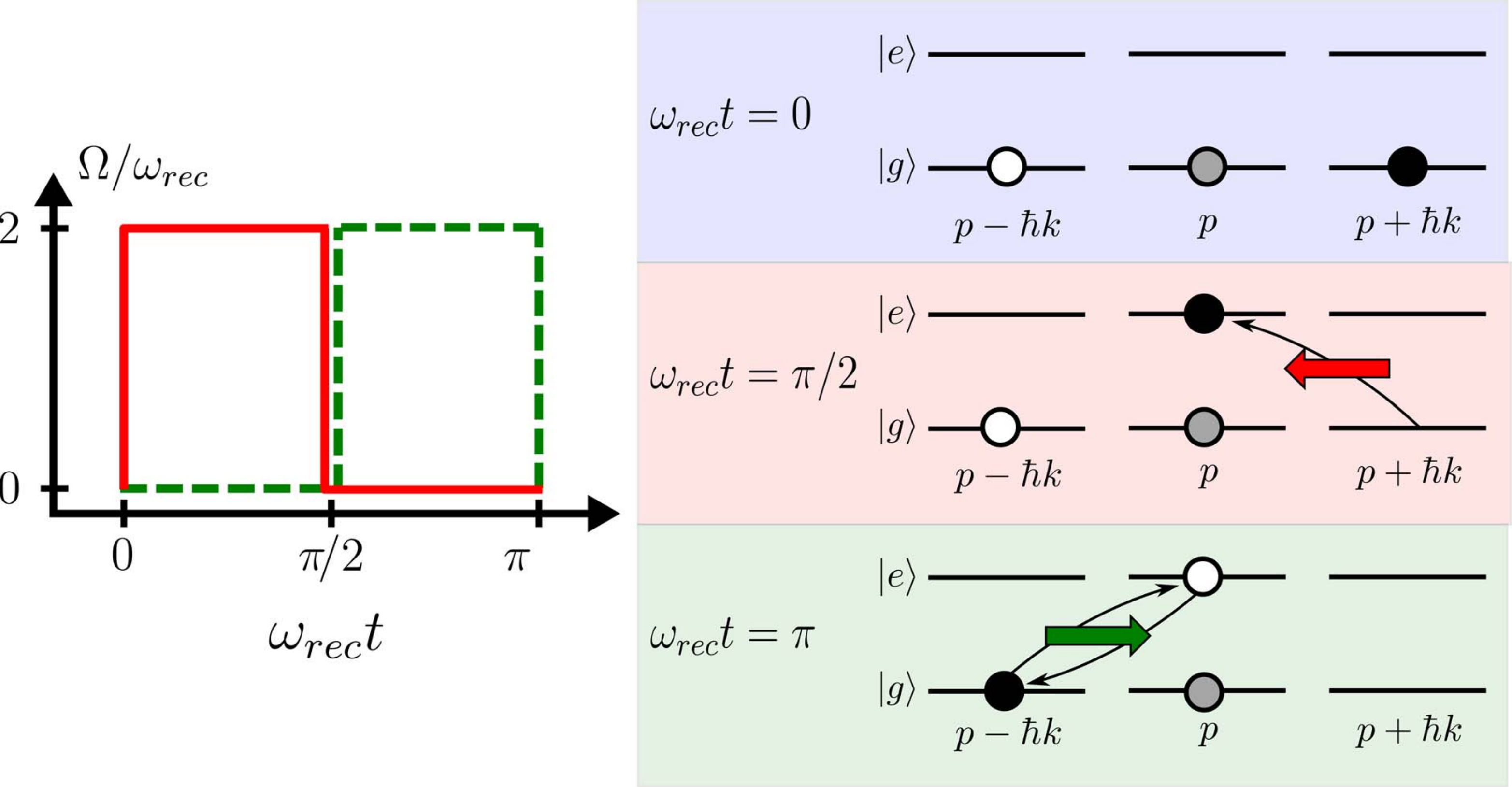}
 	\caption{Left: Pulse sequence, a $\pi$-pulse coming from the right (red) followed  by a $\pi$-pulse coming from the left (green). Right: basic idea of PMD maximum increase. The first $\pi$-pulse transfers one atom from $|p + \hbar k, g \rangle$ to $|p,e \rangle$ without affecting the atom already in state $| p,g \rangle$, thereby increasing the total number of particles in $|p \rangle$  by a factor 2. Trying to add a third particle in the same momentum $| p \rangle$ cell, by applying a second $\pi$-pulse counter-propagating, simply swaps the particles in each state with no gain in $|p\rangle$ population. $\omega_{\rm rec}=\hbar k^2/2m$ and $\Omega$ are the recoil and Rabi frequencies respectively.
}
 	\label{fig:Schema}
 \end{figure}

 \begin{figure}[h!]
 	\includegraphics[width=\columnwidth,height=1.1\columnwidth]{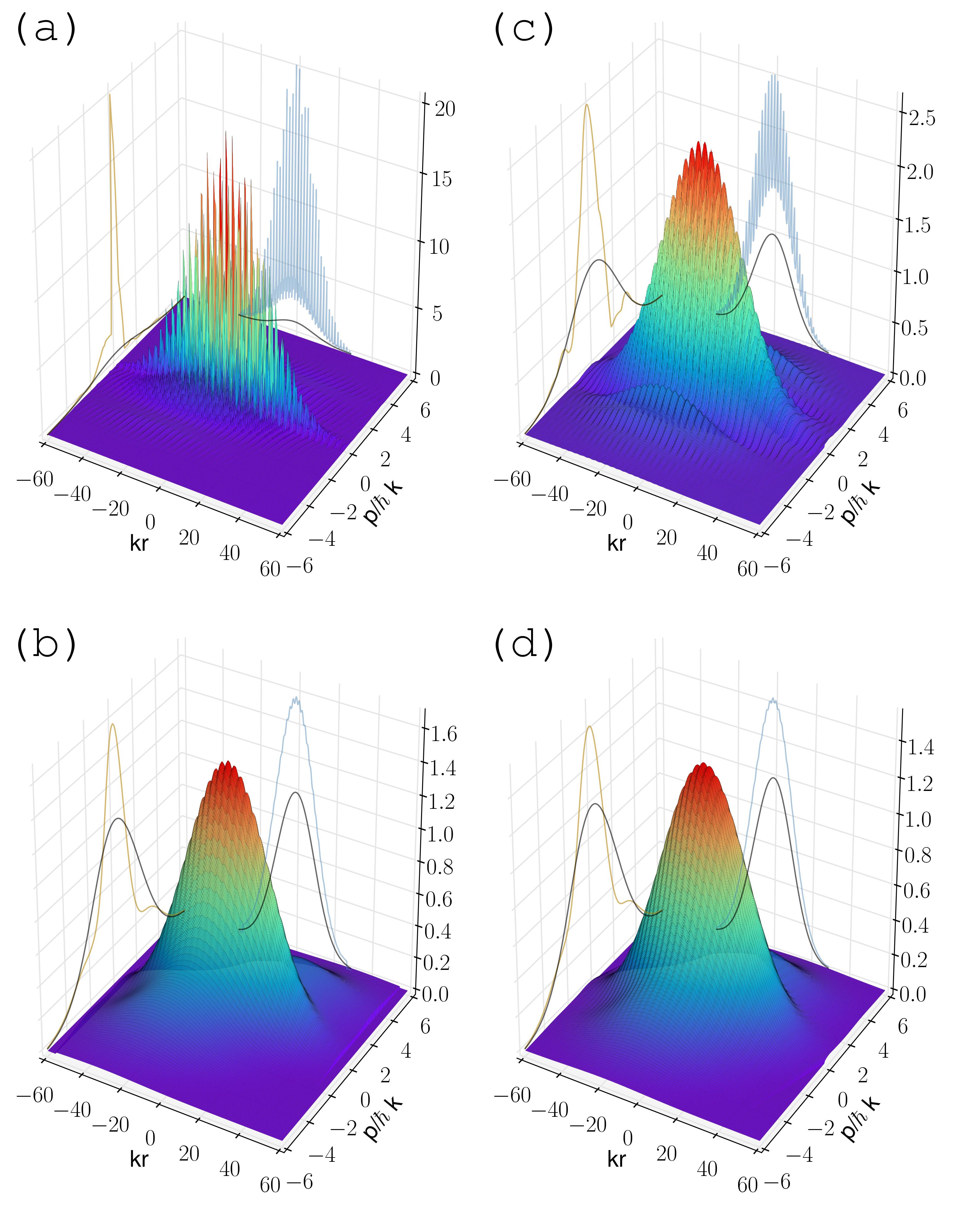}
 	\caption{PMD evolution starting with an initial Gaussian PMD (normalized to a maximum of $1$ and represented by black lines on the projections).
	 The histogram of the position momentum semi-classical evolution for cell size of $1/(5k)$ in position and $\hbar k/10$ in momentum (a) and smoothed distribution (b) as well as	 the total (ground plus excited states) Wigner (c) and Husimi (d) functions  are shown after a
pair of  $\pi$-pulses (left-right)
with Rabi frequency  $2\omega_{\rm rec}$ and pulses detuning  $-2\omega_{\rm rec}$.}
 	\label{fig:Wigner}
 \end{figure}

The basic physical mechanism and maximum gain of PMD can be  understood using an ensemble of non-interacting two-level atoms (with ground $|g\rangle$ and excited $|e\rangle$ internal states) and momentum states $|\bm p\rangle$. Because the atoms  do not interact with each other and do not undergo spontaneous emission, the one-particle Hamiltonian where the fields are classical is sufficient to describe the dynamics (see Supplemental Material(SM), Eq.(2) \cite{SM}). We ran several simulations based on various optical schemes, including bichromatic fields, rapid adiabatic transfers and $\pi$-pulses. In all cases, we found the same limitations on the gain of the PMD maximum. The underlying reasons can be understood with the example sketched in  Fig. \ref{fig:Schema}. It shows how a light pulse (with Doppler detuning and Rabi frequency $\Omega$ wisely adjusted to address a narrow line recoil transition) may bring two atoms in the same momentum state $|\bm p\rangle$ while the internal state of the displaced atom is changed. Any attempt to increase further the population of $|\bm p\rangle$ is vain because the rates of absorption and stimulated emission are equal which prevents to increase the population in $|\bm p\rangle$. This qualitatively explains the limited gain in position-momentum PSD gain by a factor 2 for 2-level atoms. 
 
We now confirm this limitation by accurate calculations for two pulses in one dimension as depicted in Fig. \ref{fig:Schema}. The classical evolution and the quantum evolution of an initial two-dimensional (thermal) Gaussian PMD in  $(r,p)$ are given in Fig. \ref{fig:Wigner}. The quantum evolution is based on the density matrix master equation $\hat{\rho}(r,p,t)$ (SM, Eq.(12)\cite{SM}) and the Wigner function $W(r,p,t)$ (SM, Eq.(13-15) \cite{SM}). 
 The semi-classical evolution makes use of Newton's equation of motion with a force (SM, Eq.(21) \cite{SM}) resulting from the Ehrenfest's theorem and Bloch equations (SM, Eq.(20) \cite{SM}) using the $\hbar k \rightarrow 0$ limit of the Wigner quantum evolution (see SM,\cite{SM}).
The evolution of the semi-classical PMD was calculated with a billion of test particles. The final distribution corresponds to the number of atoms in a position-momentum cell whose size has been arbitrarily chosen as $1/(5k)$ in position and $\hbar k/10$ in momentum. In these conditions, the maximum of the semi-classical PMD is subject to a large gain (factor 20), which significantly overcomes the quantum approaches where the maximum gain of the Wigner PMD reaches 2.5.
The semi-classical approach should indeed be handled with precaution to predict the PMD evolution.
When spontaneous emission is present, the collapse of the atomic wavepacket \cite{steuernagel1996spontaneous} smooths out the evolution on a time scale longer than the spontaneous emission time. Therefore the internal variables relax fast enough and follow quasi-adiabatically the slower external motion; so the evolution of the Wigner distribution is reduced to the semi-classical one as demonstrated in the SM \cite{SM}. On the contrary, without spontaneous emission, correlations may appear between internal and external variables \cite{dalibard1985atomic} invalidating the semi-classical approach.  

The physical relevance of the previous calculations have to be discussed in the light of the position-momentum uncertainty principle because both the quantum and semi-classical distributions exhibit structures smaller than the minimum uncertainty. This problem is often present in the distributions processed in cooling or brightening studies \cite{2015PhRvL.114d3002C,RevModPhys.89.041001,2018NJPh...20b3021N,2018arXiv180504452G, Bartolotta2018}.
This issue can be solved by performing a convolution of the PMD with a Gaussian function corresponding to the Heisenberg limit  $\sigma_r \sigma_p = \hbar/2$, which gives the smoothed coarse grained distributions  shown in Fig. \ref{fig:Wigner}(b,d), where we chose $ k \sigma_r = \frac{\sigma_p}{ \hbar k } = \frac{1}{\sqrt{2}}$.
Applied to a Wigner function, we obtain the so-called $Q(r,p,t)$ Husimi distribution which is the optimal probability distribution for joint position and momentum measurement \cite{curtright2014concise}.
 The effect is  quite striking since, in our example, the classical and quantum smoothed PMD are very similar (but still different) and both indicate a maximum gain of 2. The similitude may depend on the specificities of our toy model. Other protocols could give rise to far more significant differences. Indeed, even with a smoothing post-procedure, the semi-classical evolution should fail at the time when particles initially in the ground state and contained in an Heisenberg-bounded PSD region undergo different forces (or Rabi frequencies).

In order to  precisely understand the role of the interplay between internal and position-momentum degrees, we now adopt an analysis relying on the density matrix $\hat \rho$. For this purpose, we use the quantum PSD as a quantity linked to the entropy $S$ (per particles and per unit of $k_B$) through to the Boltzmann's formula
\begin{equation}
 S = -\ln D, \label{eq_PSD_entropy}
\end{equation} where $D$ defines the PSD quantitatively. This definition is similar to the Sackur-Tetrode formula $S = -\ln D +\frac{5}{2}$ that gives the thermal classical PSD used by the ultracold atoms community (the number of particles contained in a de Broglie's wavelength sized box reaches unity when quantum degeneration is reached). 
We first consider the Von Neuman entropy $S_{\rm VN}=- \mathrm{Tr}[ \hat{\rho} \ln(\hat{\rho}) ]=- \sum_i \lambda_i \ln(\lambda_i)$ where $\lambda_i$ are the eigenvalues of the single particle density matrix $\hat{\rho}$. These eigenstates generally do not correspond to physical observables $|i\rangle$ as the energy eigenstates for example. So other quantities are commonly used, such as the informational Shanon entropy $S_{\rm{Sh}} = -\,\sum_i p_i \ln \,p_i$ where $p_i = \langle i|  \hat \rho |i\rangle $ is the population of the $i^\mathrm{th}$  eigenstate.   Consequently, we define $D_{\rm VN}$ and $D_{\rm{Sh}}$ from Eq.(\ref{eq_PSD_entropy}).
These particular cases belong to two distinct and general categories: eigenvalue-based (or spectral) entropy and population-based (or informational) entropy. The first kind is independent of the representation basis and thus invariant under Hamiltonian evolution while the second kind depends on the representation and consequently is likely to change over time. 
In these conditions, one can wonder whether a quantum entropy can decrease or not. To answer this question, we reconsider the evolution during the pair of  $\pi$-pulses that gave rise to the PMD in Fig. \ref{fig:Wigner}. However, in order to calculate $D_{\rm{Sh}}$ and $D_{\rm{VN}}$ more easily, we now assume that the atoms are initially fully delocalized in position, which implies that the initial density matrix is Gaussian diagonal when expressed in $|p\rangle$ basis. We checked that this small modification had almost no effect on the evolution of the gain observed from the PMD (Fig. \ref{fig:Wigner} shows that the smoothed spatial distribution was almost not affected by the time evolution). 
As expected, we see in Fig. \ref{fig:psd-comparison}.(a) that the Von Neuman entropy is invariant while the Shanon entropy is not. 
More fundamentally, an initial thermal state provides the largest possible PSD and prohibits further PSD increase \cite{1992PhRvA..46.4051K}.  Indeed, the minimum Shanon entropy is achieved by a thermal Gaussian state  \cite{vinjanampathy2016quantum} and then equals the Von-Neuman entropy. So in our case $D_{\rm{Sh}}(t) \leq D_{\rm{Sh}} (0) = D_{\rm VN}(0)$. Yet it is noticeable that, unlike the Von-Neumann PSD, the Shanon PSD can locally increase as observed in Fig. \ref{fig:psd-comparison}.(a) between  $\omega_{\rm rec} t=\pi/4$ and $\omega_{\rm rec} t=\pi/2$ when the density matrix is no more Gaussian diagonal. Thus, cooling is indeed possible if starting from non-thermal states (as the one produced at time $\omega_{\rm rec} t=\pi/4$).

\begin{figure}
	\centering
	\includegraphics[width=1\linewidth]{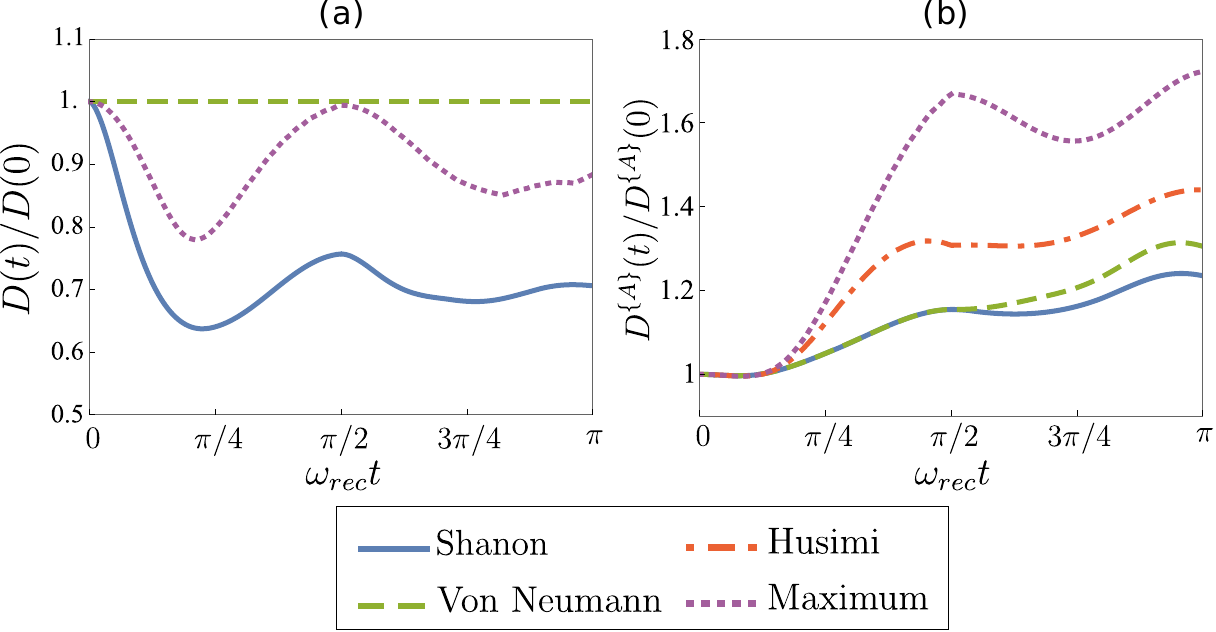}
	\caption{Evolution of several definitions of PSD, normalized to their initial value: (Husimi, $\max \left[ \hat \rho \right] $, Von Neuman  $S_{\rm VN}$ and Shanon $S_{\rm{Sh}}$ entropies) under same conditions as in Fig. \ref{fig:Wigner} but with initial full spatial delocalization. 
		a): evolution for the total (external + internal degree of freedom).
		 b): only external  degree of freedom (denoted $A$) for $\max \left[ {\hat \rho^{A} } \right] $,
  $\max \left[ Q^{\{A\} } \right]$, $D_{\rm VN}^{\{A\} }$ and $D_{\rm{Sh}}^{\{A\} }$ bounded by $2$.}		 
	\label{fig:psd-comparison}
\end{figure}

Finally, we would like to discuss the decrease of $D_{\rm{Sh}}$ and the invariance of $D_{\rm{VN}}$, which seems to contradict the results of Fig. \ref{fig:Wigner} where all the distribution maxima increase. This apparent contradiction comes from the fact that the whole density matrix we consider is composed of two subspaces: the full atomic system $AB$ ($\hat \rho = \hat \rho_{AB}$) is formed by the external degrees of freedom (part $A$) and the $M$ internal degrees of freedom $B$ (here $M=2$). As the PMD in Fig. \ref{fig:Wigner} are functions of the coordinates ($r,p$) (part $A$), it is thus more appropriate to evaluate  $S^{\{A\}}$ (or $D^{\{A\}}$), i.e. $S$ (or $D$) restricted to $A$ by using the partial trace over the internal degrees of freedom $\hat \rho_A  =Tr_B \hat \rho$ instead of $\hat \rho$. The quantity $S^{\{A\}}$ is not submitted to the constraints imposed to $S$ because entropy can be exchanged between the two subspaces. {For instance, $S_{\rm{VN}}$ verifies the subadditivity and the Araki-Leib inequality $   S_{\rm VN}^{\{AB\}} - S_{\rm VN}^{\{B\}} \leq S_{\rm VN}^{\{A\}}  \leq S_{\rm VN}^{\{AB\}} + S_{\rm VN}^{\{B\}}$ where the maximum of $S_{\rm VN}^{\{B\}}$ is $\log M$  \cite{2005PhRvA..71f3821B,bengtsson2007geometry,gemmer2009quantum,bera2016universal}}. Using Eq.(\ref{eq_PSD_entropy}), we thus find the fundamental inequality
\begin{equation}
\frac{1}{M} D^{\{AB\}}  \leq D^{\{A\}} \leq M D^{\{AB\}}
\label{eq_triangle_inequality} 
\end{equation}
that bounds the PSD evolution. The gain limit of $M$ is a fundamental result of our study. This latter also holds for $S_{\rm{Sh}}^{\{A\}}$ and consequently  $D_{\rm{Sh}}^{\{A\}}$ can only increase by a factor $M$ for an initial thermal state because  $D_{\rm{Sh}}^{\{A\}}\leq D_{\rm{VN}}^{\{A\}}$ both quantities being equal for an initial diagonal (or thermal) state.
 As discussed in the SM \cite{SM}, this is general and can be extended to other PSD definitions based on entropy, functions or maximum of PMD that are all bounded by the same factor $M$. This is consistent with our numerical results in Fig. \ref{fig:psd-comparison} showing the evolution of the quantities
 $\max \left[ {\hat \rho^{A} } \right] $,
  $\max \left[ Q^{\{A\} } \right]$, $S_{\rm VN}^{\{A\} }$ and $S_{\rm{Sh}}^{\{A\} }$ (SM, Eq.(37-38) \cite{SM}).
{As an important precaution, we mention that using pseudo phase space density definitions, as the ones filtering a specific state (such as for the ground state only $S_{\rm{Sh}}^{(g)}$, cf. SM Eq.(39) \cite{SM}), it is possible to find larger increase than a factor 2.

In conclusion, in absence of spontaneous emission and using classical laser fields, we have shown that a quantum description is more reliable than a semi-classical description of the atomic motion which can lead to large errors. We have also shown that the total eigenvalues-based PSD can not increase. This conclusion can be extended to informational population-based PSD  ($ \max \left[ \hat \rho \right]$, $S_{\rm{Sh}}$ entropy or $\max\left[ Q \right]$) when the initial state is a diagonal state. Still, a sample initially prepared in a thermal state and thereby without quantum correlation can exhibit a gain of the PMD maximum or PSD up to the number $M$ of internal states (or ultimately $M^2$ if initial correlations exist in the initial state, see SM \cite{SM}). 
The direct and fundamental consequence of this analysis, holding for any kind of free particles or particles in time-dependent trapping potential is that cooling mechanisms based on coherent field momentum transfer without spontaneous emission (such as adiabatic passages, bichromatic, $\pi$-pulses \cite{vitanov2017stimulated, 2018NJPh...20b3021N, 2018arXiv180504452G,PhysRevA.54.R1773,PhysRevA.85.033422,RevModPhys.89.041001}) have a limited efficiency and could only lead to a position-momentum PSD gain of $M$. This is still of interest for studies that need more particles in a same phase space area regardless of the internal distribution (for laser manipulation, detection, collisional studies, ...).
However, increasing the full PSD is impossible, in other words, the production under coherent fields of all particles in the same internal state with a larger PSD than the initial one is impossible without spontaneous emission. }
An obvious way to overcome this limitation is to allow a single spontaneous emission event per particle \cite{pyshkin2016ground,2014PhRvA..89b3425J,2014PhRvA..89d3410C}  because the third ancilla spontaneous emission space has almost an infinite dimension to extract entropy (see  	\cite{1986PhRvA..34.4728C,1988AnPhy.186..381P,2012CoTPh..57..209Y,2008CSF....37..835Y,2009OptCo.282.2642X,2010IJTP...49..276A,2012PhyA..391..401Z,2013IJTP...52.1122K,1992PhRvA..46.1438V,2008PhRvA..77f1401M,2007PhRvB..75u4304R}).
A second option for cooling is to create entanglement between particles and the light field \cite{beige2005cooling,vacanti2009cooling} or by using non statistical methods such as 	informational cooling (stochastic cooling being one famous example)  \cite{1998PhRvA.58.4757R,2001PhRvA..64f3410B} or cavity cooling  \cite{2000PhRvL..84.3787V,2005PhRvA..71f3821B,2000JMOp...47.2741G,2000PhRvL..84.3787V,murr2006large,2013AnPhy.334..272C,beige2005cooling}.
A final alternative would be to use non-classical quantum fields. Because absorption or stimulated emission rates are not equivalent anymore (with the simple example of Fock states), the last step sketched in Fig. \ref{fig:Schema} would allow one to put more atoms at the same phase space location \cite{dalibardcoolegeFrance2015}. In other words, when the optical field is no longer considered as a parameter, the total system is now composed of 3 sub-systems (external, internal degrees of freedom and quantized field). Our previous demonstrations could then be applied: the (external) PSD can be increased by the number of available micro-states in the other (internal and field) spaces. If the latter are sufficiently large, there is a priori no theoretical limit on cooling even without spontaneous emission \cite{2008PhRvA..77f1401M,2015PhRvL.114d3002C,2015JOSAB..32B..75C,RevModPhys.89.041001,2000PhRvL..84.3787V,2005PhRvA..71f3821B,murr2006large}.

Acknowledgment: The authors thank P. Cheinet for their valuable advice. This work was supported by ANR MolSisCool, ANR HREELM, Dim Nano-K CPMV, CEFIPRA No. 5404-1, LabEx PALM ExciMol and ATERSIIQ (ANR-10-LABX-0039-PALM).

\title{Supplemental Material for \\ Phase space density limitation in laser cooling without spontaneous emission}
\author{Thierry Chaneli\`{e}re, Daniel Comparat and Hans Lignier}
\affiliation{Laboratoire Aim\'{e} Cotton, CNRS, Univ. Paris-Sud, ENS Paris Saclay, Universit\'e Paris-Saclay, B\^{a}t. 505, 91405 Orsay, France}

\date{\today}
\maketitle
\onecolumngrid

\section{Non-relativistic Hamiltonian of non-interacting particles}

We here recall the equations of motion for laser cooling of atoms. The reader can refer to textbooks such as 
\cite{CDG2}.

\subsection{Quantized or (semi-)classical hamiltonian}

We  here study the quantum Hamiltonian $\hat H$ of a two generic levels $|1\rangle$ and $|2\rangle$ (representing the ground $|g\rangle$ and the excited $|e\rangle$  states in \cite{main_art}) of a particle (mass $m$) under the effect of electromagnetic fields. The generalisation to $M$ level system is straightforward but will not be detailed for the sake of simplicity. We separate the "motional" (or trapping) fields that do not couple $|1\rangle$ and $|2\rangle$, such as trapping potential $V_1,V_2$  produced for example by magnetic coils, magnets or electrodes through Zeeman  ($-\hat{\bm \mu} . \bm B$) or   Stark effect ($-\hat{\bm d}.\bm E$), and the laser fields $\hat{\bm E}$  that do couple $|1\rangle$ and $|2\rangle$. 

For $N$  non-interacting particles the full hamiltonian can be written as
$\hat H=  \sum_{i=1}^N \hat H^{(i)} + \hat H_{\rm field} + \sum_{i=1}^N \hat H_{\rm int, field}^{(i)} $, where $\hat H^{(i)}$ is the hamiltonian $\displaystyle \frac{\bm \hat p_i^2}{2 m} + 
V_1(\hat{\bm r}_i,t) |1\rangle\langle 1| + V_2(\hat{\bm r}_i,t) |2\rangle \langle 2|$ for the position and momentum $\bm p_i,\bm r_i$ of the $i^\mathrm{th}$ particle. The trapping field is arbitrary but the simplest case corresponds to harmonic traps: $V_i = E_i +\frac{1}{2} m \omega_i \bm r^2$. A base of the Hilbert space will be an ensemble of states $\displaystyle \bigotimes_{i=1}^{N} |\bm p_i,1\ {\rm or} \ 2\rangle_i \otimes | \Uppi_{\bm{k} \sigma}  n_{\bm{k} \sigma} \rangle$ when using the Fock notation for the field.
We treat the $N$ particles as totally independent  and use the 
density matrix formalism (written as $\hat \rho$) to describe the system of $N$ identical particles as a statistical ensemble. The external field is common to the $N$ atoms and this can automatically generate entanglement between the atoms or collective behaviour  that can indeed lead to cooling \cite{beige2005cooling,vacanti2009cooling}. 
As explained in the article, this is not our interest here and we shall study only the single particle case.
In the dipolar approximation and neglecting the 
Roentgen term, despites the fact  that it can create surprising radiation forces  on the atoms \cite{barnett2017vacuum,2017arXiv170401835S},
the Hamiltonian for a single particle reads as:
\begin{equation}
\hat H =  \frac{\hat{\bm p}^2}{2 m} + 
V_1(\hat{\bm r},t) |1\rangle\langle 1| + V_2(\hat{\bm r},t) |2\rangle \langle 2| \label{eq_base_at}
 -  \bm d. \hat{\bm E} (\hat{\bm r} ,t) (   |2\rangle \langle 1| + |1\rangle \langle 2| ) +
\sum_{\bm{k} \sigma} \hbar \omega_k \left( \hat a^\dag_{\bm{k} \sigma} \hat a_{\bm{k} \sigma} + 1/2 \right) 
\end{equation}
where $\bm d$ is the transition dipole element  (assumed to be real $\bm d = \langle 2 | q \hat {\bm r}| 1 \rangle $) and
 $\hat{\bm E} (\bm r ,t)$ is a quantized real field. For instance for a single plane wave field (in a volume $L^3$)
$\displaystyle	\hat{\bm{E}} (\bm{r},t)  = \sum_{\bm{k},\sigma} i \sqrt{\frac{ \hbar  \omega_k}{2  \epsilon_0 L^3} } \left(  \hat a_{\bm{k} \sigma} e^{- i \omega_k t}  \bm{\epsilon}_{\bm{k}\sigma} {\rm e}^{i \bm{k}.\bm{r} }  - \hat a_{\bm{k} \sigma}^\dag  e^{ i \omega_k t}  \bm{\epsilon}_{\bm{k}\sigma}^\ast {\rm e}^{-i \bm{k}.\bm{r} } \right)$. 


The initial state is uncorrelated and  density operator can be written as an atomic (external and internal degrees of freedom) and a field part as
$\displaystyle \hat \rho = \hat{\rho}_{\rm at} \otimes \hat{\rho}_{\rm field} =  \hat{\rho}_{\rm ext} \otimes  \hat{\rho}_{\rm int} \otimes  \hat{\rho}_{\rm field}  $.

In the semi-classical approximation, we would like to replace the field operators (denoted with the hat $ \hat{}  $ ) by their classical expectation values, namely
$\hat a_{\bm{k} \sigma}$ and $\hat a_{\bm{k} \sigma}^\dagger$ by c-numbers
$ a_{\bm{k} \sigma}$ and $ a_{\bm{k} \sigma}^*$, such as 
	$ \hat{\bm E} (\hat{\bm r} ,t)$ by $   \bm E (\hat{\bm r} ,t)$
  becomes in the Hamiltonian
\begin{equation}
\hat H =  \frac{\hat{\bm p}^2}{2 m} + 
E_1(\hat{\bm r},t) |1\rangle\langle 1| + E_2(\hat{\bm r},t) |2\rangle \langle 2|
 - \bm d. \bm E (\hat{\bm r} ,t) (   |2\rangle \langle 1| + |1\rangle \langle 2| ) \nonumber \label{Hamil}
\end{equation}

\subsection{Classical fields}

This can be done, 
by using coherent states $|\alpha \rangle$,
that are eigenstates of the annihilation operator $\hat a$: $\hat{a}|\alpha\rangle=\alpha|\alpha\rangle$, by using the unitary transformation under the operator $\hat U =  \hat {\cal D}(\alpha_\lambda e^{-i\omega_\lambda } )^\dag $ and neglecting the quantum field that now describes  spontaneous emission only \cite{mollow1975pure,Cohen-Tannoudji1997,dalibardcoolegeFrance2015}.


Therefore, in the following
we assume  to have classical laser fields with different frequencies $\omega_{\rm L}$, wave-vectors ${\bm k}_{\rm L}$ or temporal phase $\Phi_{\rm L}(t)$:
${\bm E}(\hat{\bm r},t) = 
{\bm E'}(\hat{\bm r},t)+ {\bm E'}^\dag(\hat{\bm r},t) =
\frac{1}{2}\sum_{\rm L}    \left[ {\bm E}_{\rm L}(t) e^{i(  {\bm k}_{\rm L}. {\hat{\bm r}} - \omega_{\rm L} t   - \Phi_{\rm L}(t)
	)} +  {\bm E}_{\rm L}^\ast(t) e^{-i(  {\bm k}_{\rm L}. {\hat{\bm r}}  - \omega_{\rm L} t   - \Phi_{\rm L}(t)
	)}  \right] $.
The rotating wave approximation leads to

\begin{equation}
\hat H =  \frac{\bm \hat {\bm p}^2}{2 m} + 
V_1(\hat {\bm r},t) |1\rangle\langle 1| + V_2(\hat {\bm r},t) |2\rangle \langle 2| 
 -  \bm d .\bm E' (\hat {\bm r} ,t)  |2\rangle \langle 1| - \bm d .  \bm E'^\dag (\hat {\bm r} ,t)  |1\rangle \langle 2|) \label{hamiltonianRWA}
\end{equation}

We will now use this Hamiltonian to describe the evolution.
In matrix notation with the  $|1,2\rangle$ basis,  the Hamiltonian (\ref{hamiltonianRWA}) becomes
$\hat H = \begin{pmatrix}
\hat H_1 & \hat V^\dag \\ 
\hat V & 	\hat H_2
\end{pmatrix} $
where the coupling term is  $\displaystyle \hat V = -  \bm d .\bm E' (\hat r ,t) =
- \frac{\bm d}{2}\sum_{\rm L}    {\bm E}_{\rm L}(t) e^{i(  {\bm k}_{\rm L}. {\hat{\bm r}} - \omega_{\rm L} t  - \Phi_{\rm L}(t) )}		= \sum_{\rm L}	\hat V_{\rm L}	$.

\subsubsection{Density matrix}

The time evolution 
$\displaystyle i \hbar \frac{\partial \hat \rho}{\partial t} = \hat H \hat \rho - \hat \rho \hat H$ leads to:
\begin{widetext}
	$ \begin{pmatrix}
	\frac{\partial \hat \rho_{11} }{\partial t}	 & 	\frac{\partial \hat \rho_{12} }{\partial t} \\ 
	\frac{\partial \hat \rho_{21} }{\partial t} & 	\frac{\partial \hat \rho_{22} }{\partial t}
	\end{pmatrix} = \frac{1}{i\hbar}
	\begin{pmatrix}
	[\hat H_1,\hat \rho_{11}]+ \hat V^\dag \hat \rho_{21} - \hat \rho_{12 } \hat V   & [\hat p^2/2m,\hat \rho_{12}] + \hat V_1 \hat \rho_{12} - \hat \rho_{12} \hat V_2 + \hat V^\dag \hat \rho_{22} - \hat \rho_{11} \hat V^\dag\\ 
	[\hat p^2/2m,\hat \rho_{21}] + \hat V_2 \hat \rho_{21} - \hat \rho_{21} \hat V_1 + \hat V \hat \rho_{11} - \hat \rho_{22} \hat V & 	[\hat H_2,\hat \rho_{22}]
	+ \hat V\hat \rho_{12} - \hat \rho_{21 } \hat V^\dag
	\end{pmatrix}
	$
\end{widetext}

\subsubsection{Wigner functions}
\onecolumngrid

The Wigner-Weyl transform of this equation gives the time evolution of the Wigner function defined as

	\begin{equation}
	W(\bm r,\bm  p,t) = \frac{1}{h^3}
	\int \langle  \bm  p - \bm  p'/2 | \hat \rho (\hat{\bm  r},\hat{ \bm p}, t) | \bm  p + \bm p'/2 \rangle e^{-i \bm r.\bm  p'/\hbar} d\bm  p' 
	\label{def_Wigner} 
	\end{equation}

through the so-called Moyal bracket, governed by

\begin{equation}	\frac{\partial W}{\partial t} =  \frac{1}{i \hbar} \left(H \star W - W \star H \right) \end{equation}

The  $\star$-product
can
be evaluated using the convenient formula
\cite{curtright2014concise} for any generic function $\rho_{1,2}(r,p)$
\begin{eqnarray*}
	(\rho_1 \star 	\rho_2) (r,p) &= & 	\rho_1(r+i\frac{\hbar}{2} \frac{\partial}{\partial p},p-i\frac{\hbar}{2} \frac{\partial}{\partial r}) 	\rho_2(r,p) \\
	(\rho_2 \star 	\rho_1) (r,p) &= & 	\rho_2(r-i\frac{\hbar}{2} \frac{\partial}{\partial p},p+i\frac{\hbar}{2} \frac{\partial}{\partial r}) 	\rho_1(r,p) 
\end{eqnarray*}
that we have restricted to a one dimensional motion for simplicity.

Therefore,
when no  $\hat r$, $\hat p$ product are present in $\hat \rho = \rho(\hat r, \hat p)$,
the  Wigner(-Weyl) transform $W_{\hat \rho}(r,p;t)$  is  the unmodified classical observable expression $\rho(r,p)$. An important example is a conventional Hamiltonian, $\displaystyle \hat H = \hat p^2/2m +V(\hat r,t)$, for which the transition from classical
mechanics  is the straightforward quantization: $\displaystyle W_{\hat H}(r,p;t) = H(r,p;t) = p^2/2m +V(r,t)$.

The expressions containing  $ e^{i  {\bm k}_{\rm L}. {\hat{\bm r}} }$
can be expanded by
using exponential (Taylor) series  that indicates	
$e^{i  { k}_{\rm L} \left(
	r \pm \frac{i  \hbar}{2}   \frac{\partial }{\partial  p}
	\right) } f(r,p,t) = 
e^{i   k_{\rm L} r} f(r,p  \mp   \hbar   k_{\rm L} /2   ,t)
$.
and finaly using $\displaystyle \hbar \Upomega_{\rm L}(r,t) = 
\bm d. {\bm E}_{\rm L} e^{ i ( {\bm k}_{\rm L}. {\bm r}  - \omega_{\rm L} t  - \Phi_{\rm L}(t) )} $,
 we obtain:
	{\footnotesize
	\begin{eqnarray}
	\left[ \frac{\partial   }{\partial t}	 
	+\frac{ p}{m} \frac{\partial }{\partial  r}   
	- 
	\frac{1}{i\hbar}	[V_1(r+i\frac{\hbar}{2} \partial_p) - V_1(r-i\frac{\hbar}{2} \partial_p) ] \right]W_{11}(r,p,t) &= &  
	-
	\frac{1}{2 i} \sum_{\rm L} ( \Upomega_L^* (r,t)
	W_{21} (r,p+\frac{\hbar k_{\rm L}}{2},t) - \Upomega_L (r,t) W_{12} (r,p+\frac{\hbar k_{\rm L}}{2},t)
	) \label{Wigner_eq_brute_1}
 	\\
	\left[	\frac{\partial   }{\partial t}+\frac{ p}{m}  \frac{\partial }{\partial  r} 
	-
	\frac{1}{i\hbar}	[V_1(r+i\frac{\hbar}{2} \partial_p) - V_2(r-i\frac{\hbar}{2} \partial_p) ] \right] W_{12}(r,p,t) & = &
	-
	\frac{1}{2 i} \sum_{\rm L} \Upomega_L^* (r,t) (
	W_{2 2} (r,p+\frac{\hbar k_{\rm L}}{2},t) - W_{11} (r,p-\frac{\hbar k_{\rm L}}{2},t)
	)
	\label{Wigner_eq_brute_2}
	\\
	\left[	\frac{\partial   }{\partial t}  +\frac{ p}{m}  \frac{\partial }{\partial  r} 
	- 
	\frac{1}{i\hbar}	[V_2(r+i\frac{\hbar}{2} \partial_p) - V_1(r-i\frac{\hbar}{2} \partial_p) ] \right] W_{21}(r,p,t) & = &
	-
	\frac{1}{2 i} \sum_{\rm L} \Upomega_L (r,t) (
	W_{11} (r,p-\frac{\hbar k_{\rm L}}{2},t) - W_{22} (r,p+\frac{\hbar k_{\rm L}}{2},t)
	)
	\label{Wigner_eq_brute_3}
	\\	
	\left[	\frac{\partial  }{\partial t}
	+\frac{ p}{m}  \frac{\partial }{\partial  r}   
	-	\frac{1}{i\hbar}	[V_2(r+i\frac{\hbar}{2} \partial_p) - V_2(r-i\frac{\hbar}{2} \partial_p) ] \right]   W_{22} (r,p,t)   &=  &
	- \frac{1}{2 i} \sum_{\rm L} ( \Upomega_L (r,t)
	W_{12} (r,p-\frac{\hbar k_{\rm L}}{2},t) - \Upomega_L^* (r,t) W_{21} (r,p-\frac{\hbar k_{\rm L}}{2},t)
	) 
	\label{Wigner_eq_brute_4}
	\end{eqnarray}
	}
For completeness, we mention that a (1D) spontaneous emission rate $\Gamma$ can be added if needed, by including the terms \cite{dalibard1985atomic,1991JOSAB...8.1341Y}.

\begin{eqnarray*}
	\left.	\frac{\partial W_{11}}{ \partial t}	\right|_{\rm spon}  & =& \Gamma \int_{-p_r}^{p_r} \Theta(p') W_{22}(r,p+p') d p' \\
	\left.	\frac{\partial W_{11}}{ \partial t} 	\right|_{\rm spon}  & = &  - \frac{\Gamma}{2}  W_{12}(r,p)\\
	\left.	\frac{\partial W_{21}}{ \partial t} 	\right|_{\rm spon}  & = &  - \frac{\Gamma}{2}  W_{21}(r,p) \\	
	\left.		\frac{\partial W_{22}}{ \partial t} 	\right|_{\rm spon}  & = &  - \Gamma  W_{22}(r,p)
\end{eqnarray*}
where $\Theta(p')$ is the probability density distribution 
for 
the projection of spontaneous emission $\displaystyle \Theta(p')=  \frac{3}{8 p_r}\left(1 +\frac{p'^2}{p_r^2} \right)$ for a dipolar radiation pattern) on the atomic recoil momentum  for $ p_r = \hbar k$.

Equation of motion of the Husimi distribution can be derived \cite{lee1995theory,takahashi1989distribution,martens2011quantum,wyatt2006quantum,chattaraj2016quantum}
         and  present non-zero second term of the Liouville equation (similar to Eqs.(\ref{Wigner_eq_brute_1})-(\ref{Wigner_eq_brute_4}))

	\subsection{Connection with Liouville equation}
	
In the absence of light fields, Taylor series expansion indicates that the evolution of the diagonal terms $W_{ii}$ is  given by:
	\begin{equation*}
\frac{D W_{ii}}{D t} = 	\frac{\partial W_{ii}}{\partial t} + \frac{ p}{m} \cdot \frac{\partial W_{ii}}{\partial  r} - \frac{\partial V_i}{\partial  r}\cdot   \frac{\partial W_{ii}}{\partial  p}  = 
	 \sum_{s\geq 1} \hbar^{2s} \frac{2^{-2s}}{(2s+1)!} \frac{\partial^{2s+1} V_i}{\partial r^{2s+1} }
	\frac{\partial^{2s+1} W_{ii}}{\partial p^{2s+1} }
	\label{Wigner_evolution}
	\end{equation*}
	We recover the 
Liouville's equation, $\displaystyle \frac{D W_{ii}}{D t} = 0$, under the influence of the
	potential $V$, but only 
	for a quadratic potential $V_i(r,t) = a(t) + b(t) r +c(t) r^2$.
	However, when higher derivatives of $V_i(r)$ are present, additional terms will give rise to diffusion and the quantum Wigner function
	gradually deviates from the corresponding classical phase space probability density. So a non-harmonic potential is a clear way to modify the Wigner phase space density. This argument also applies to the Husimi function.

\subsection{Interaction picture: free evolution}

The evolution  of $H_1(t)$ is given by the unitary time evolution operator 	$\hat U_1 (t) = e^{-i \int \hat H_1(t)/\hbar }$. 
In matrix notation, the		evolution operator is
$\hat U_0 = \begin{pmatrix}
\hat U_1 & 0 \\ 
0 & 	\hat U_2
\end{pmatrix} $.
The interaction picture consists in defining 
a new density matrix
$\displaystyle \hat \rho^I (t) = {\hat U_0}^\dag (t) \hat \rho (t) {\hat U_0} (t)$, which 
evolves under the
modified Hamiltonian 
$\displaystyle \hat H^I  = {\hat U_0}^\dag  \hat H {\hat U_0}  + i  \hbar \frac{d {\hat U_0}^\dag}{d t}   {\hat U_0}  =
\begin{pmatrix}
0 &{{\hat V}^I} {}^{\dag} \\ 
{{\hat V} {}^I} & 0
\end{pmatrix}$ where $\hat V^I = {{\hat U}_2}^\dag \hat V \hat U_1$.

Because several laser frequencies are possibly present, the interaction picture is more appropriate than the Bloch rotating frame. The latter would imply to choose one laser frequency as a reference. The interaction picture removes this arbitrariness.

\subsubsection{Density matrix}
Using the momentum representation, where $\hat r$ acts as $i \hbar \partial_p$  on $ \psi(p) = \langle p | \psi\rangle$, we have 
$\displaystyle   e^{i k \hat r  } | p \rangle = | p + \hbar k \rangle $
We find
\begin{eqnarray}
\hat V^I  | p \rangle &= & - \frac{1}{2} \sum_{\rm L} | p + \hbar k_L \rangle \Omega_{\rm L}  e^{-i ( \delta_{\rm L}^{p+} t )} \label{Vinter} \\
\delta_{\rm L}^{p\pm} & =&  \omega_{\rm L} - (E_2-E_1)/\hbar -  \frac{k_{\rm L}}{m}(p   \pm \hbar k_{\rm L} /2 )  
\end{eqnarray}
where  $\hbar \Omega_{\rm L}(t) = \bm d. {\bm E}_{\rm L} e^{ - \Phi_{\rm L}(t)}$ and 
$\delta_{\rm L}^{p\pm} =  \delta_{\rm L}^0 + \delta_{\rm L}^{\rm D}(p) \pm \delta_{\rm L}^{\rm r}$:
The detuning $\delta_{\rm L}^0 = \omega_{\rm L} - (E_2-E_1)/\hbar $, the Doppler shift $ \delta_{\rm L}^{\rm D}(p) = -  k_{\rm L}. p /m $ and recoil frequency $ \delta_{\rm L}^{\rm r} = - \hbar k_{\rm L}^2 /2 m  $ appear naturally.

With $\hat \rho^I_{i j} = {\hat U_i}^\dag \hat \rho_{i j} {\hat U_j}$,  
the evolution reads as:

\begin{equation} \begin{pmatrix}
\frac{\partial \hat \rho_{11}^I }{\partial t}	 & 	\frac{\partial \hat \rho_{12}^I }{\partial t} \\ 
\frac{\partial \hat \rho_{21}^I }{\partial t} & 	\frac{\partial \hat \rho_{22}^I }{\partial t}
\end{pmatrix} = \frac{1}{i\hbar} \sum_{\rm L}
\begin{pmatrix}
{{\hat V} {}^I}^\dag \hat \rho_{21}^I - \hat \rho_{12 }^I \hat V^I   &  	{{\hat V} {}^I}^\dag \hat \rho_{22}^I - \hat \rho_{11}^I 	{{\hat V} {}^I}^\dag\\ 
{{\hat V} {}^I} \hat \rho_{11}^I - \hat \rho_{22}^I 	{{\hat V} {}^I}& 
{{\hat V} {}^I}\hat \rho_{12}^I - \hat \rho_{21 }^I 	{{\hat V} {}^I}^\dag
\end{pmatrix}
\end{equation}

Assuming there is no external field from now and using  $\displaystyle {\rho^I}_{i j }^{p' p} = \langle p' | \hat \rho_{i j }^I | p\rangle  = e^{i (p'^2-p^2) t/2m \hbar} e^{i (E_i-E_j) t/\hbar} \rho_{i j }^{p' p} $, the latter can be written as:

{\footnotesize
	\begin{equation}
	\begin{pmatrix}
	\frac{\partial   {\rho^I}_{11}^{p' p} }{\partial t}	 & 	\frac{\partial  {\rho^I}_{12}^{p' p} }{\partial t} \\ 
	\frac{\partial  {\rho^I}_{21}^{p' p} }{\partial t} & 	\frac{\partial  {\rho^I}_{22}^{p' p} }{\partial t}
	\end{pmatrix} = - \frac{1}{2 i} \sum_{\rm L}
	\begin{pmatrix}
	\Omega_{\rm L}^*	e^{i  \delta_{\rm L}^{p'+} t   }   {\rho^I}_{21}^{(p'+\hbar k_{\rm L})  p} -   \Omega_{\rm L} {\rho^I}_{12 }^{p' (p+\hbar k_{\rm L})}   e^{-i  \delta_{\rm L}^{p+} t   }    &  	 \Omega_{\rm L}^* e^{i  \delta_{\rm L}^{p'+} t   }   {\rho^I}_{22}^{(p'+\hbar k_{\rm L}) p} -  \Omega_{\rm L}^* {\rho^I}_{11}^{p' (p-\hbar k_{\rm L})}  e^{i  \delta_{\rm L}^{p-} t  } \\ 
	\Omega_{\rm L}	e^{-i  \delta_{\rm L}^{p'-} t   }   {\rho^I}_{11}^{(p'-\hbar k_{\rm L}) p} -  \Omega_{\rm L} {\rho^I}_{22}^{p' (p+\hbar k_{\rm L})}   e^{-i  \delta_{\rm L}^{p+} t   }  & 
	\Omega_{\rm L}	e^{-i  \delta_{\rm L}^{p'-} t  }  {\rho^I}_{12}^{(p'-\hbar k_{\rm L}) p} -  \Omega_{\rm L}^* {\rho^I}_{21 }^{p' (p-\hbar k_{\rm L})}  e^{i  \delta_{\rm L}^{p-} t   } 
	\end{pmatrix} \label{eq_density}
	\end{equation}
}

\subsubsection{Wigner function}

It is quite convenient to use
the so-called non-diagonal Wigner functions 
by defining $W_{i j}^I = W_{\hat \rho_{i j}^I}/h$ as the Wigner transform function associated to $ \hat \rho_{i j}^I = \langle i |\hat \rho^I | j\rangle$.
So
$ W_{i j}(r,p,t) = e^{i (E_j-E_i) t/\hbar} W_{i j}^I(r-p t/m,p,t)  
$ and the evolution equations become:

	\begin{eqnarray}
	\frac{\partial  W_{11}^I }{\partial t}	 
	(r,p,t) & =&   
	- \sum_{\rm L}
	\Im \left[   \Omega_L^* (r,p,t)
	W_{21}^I (r- \hbar k_{\rm L} t /2m,p+\hbar k_{\rm L}/2,t)
	\right] \label{int_wigner_a}
	\\
	\frac{\partial   W_{21}^I }{\partial t}  (r,p,t) & = &
	\frac{1}{2 i} \sum_{\rm L} \Omega_L(r,p,t) (W_{22}^I (r- \hbar k_{\rm L} t /2m,p+\hbar k_{\rm L}/2,t) -
	W_{11}^I (r+ \hbar k_{\rm L} t /2m,p-\hbar k_{\rm L}/2,t) 
	) 
	\label{int_wigner_b}
	\\	
	\frac{\partial 	W_{22}^I }{\partial t}
	(r,p,t)  & = & 
	\sum_{\rm L}
	\Im \left[   \Omega_L^*(r,p,t) W_{21}^I (r+ \hbar k_{\rm L} t /2m,p-\hbar k_{\rm L}/2,t)
	\right]
	\label{int_wigner_c}
	\end{eqnarray}

where 
\begin{equation}
\Omega_{\rm L}(r,p,t)  = \Omega_{\rm L} e^{i (k_L r   +  k_{\rm L} p t /m  -  \delta_{\rm L}^0 t   - \Phi_{\rm L}(t))	}
\label{Omega}
\end{equation}

\subsection{Single laser case (Bloch equation)}

When there is only one laser, we can define

\begin{eqnarray*}
	\tilde W_{11}^I(r,p,t) &= &   W_{11}^I (r,p,t)  \\
	\tilde W_{22}^I (r,p,t)& =&   W_{22}^I (r- \hbar k_{\rm L} t /m,p+\hbar k_{\rm L},t) \\
	\tilde W_{21}^I (r,p,t) 	&= & e^{-i (k_{\rm L} r   + \frac{ k_{\rm L} p t}{m}  -  \delta_{\rm L}^0 t  - \Phi_{\rm L} )}  W_{21}^I (r- \frac{\hbar k_{\rm L} t}{ 2m},p+ \frac{\hbar k_{\rm L}}{2},t)
\end{eqnarray*}

If we assume $\Omega_{\rm L}$ real, the evolution is governed by

\begin{eqnarray}
	\frac{\partial   }{\partial t}	 
	\frac{
		\tilde	W_{11}^I  - \tilde W_{22}^I 
	}{2} & =&   
	-  \Omega_L \Im  
	\tilde W_{21}^I   + \frac{ \hbar k_{\rm L}}{ 2 m} \frac{\partial   }{\partial r}   \tilde W_{22}^I 
	\\
	\frac{\partial   }{\partial t}  \Re \tilde W_{21}^I & = & - 	\delta_{\rm L}^{p+}   \Im \tilde W_{21}^I -\frac{ \hbar k_{\rm L}}{  2m} \frac{\partial   }{\partial r}   \Re \tilde W_{21}^I
	\\
	\frac{\partial   }{\partial t}  \tilde \Im W_{21}^I & = & 	\delta_{\rm L}^{p+}  \Re \tilde W_{21}^I
	+
	\Omega_L 	\frac{\tilde W_{11}^I  - \tilde W_{22}^I }{2 }   -\frac{ \hbar k_{\rm L}}{  2m} \frac{\partial   }{\partial r}   \Im \tilde W_{21}^I
\end{eqnarray}

We recognize the standard Bloch equations except for the term in $\displaystyle \frac{ \hbar k_{\rm L}}{  2m} \frac{\partial   }{\partial r}$.
We can thus retrieve the Bloch equations from the exact Wigner function evolution by performing series expansion in $\hbar k$. 
This approach justifies the semi-classical equation for the particles evolution that we derive from heuristic considerations.

\section{Semi-classical evolution}
\label{Semiclassical evolution}

From the quantum evolution, we can derive the semi-classical evolution of the atomic motion. The underlying assumption is that
the displacement of the atom during the internal relaxation time is very small. The internal variables follow quasi-adiabatically the external motion \cite{dalibard1985atomic}. It is then possible to separate the internal and the external degree of freedom.

The Doppler or recoil effects, or the use of the stationary state of the Bloch equation can be done with hand-waving arguments  
(see for instance in Ref. \cite{2016PhRvA..93f3410H}). Nevertheless, the Lagrangian description (individual particles are followed through time), Eulerian description and interaction picture  that freeze the motion in the Eulerian description may  lead to confusion. We will clarify this distinction.

\subsection{Definition of a force}
\label{Force_Bloch}

For simplicity, we neglect the external potentials (but they can be included in the interaction picture if needed).

In the semi-classical approach, the particle motion is classical: for a given particle initially at $\bm r(t_0)=\bm r_0$ and  $\bm v(t_0)=\bm v_0$ at time $t=t_0$ its trajectory in phase space  $\bm r(t ),\bm p(t)=m \bm v(t)$ is given 
by Newton's equation of motion 
$\displaystyle m \frac{ d \bm v}{d t} (t)= \bm F(\bm r(t),\bm v(t), t)$.

The standard way to define the force in laser cooling is by using the Ehrenfest theorem 
(see for instance \cite{cohen1990atomic,Met1}, but other methods exists \cite{romanenko2017atoms,podlecki2017radiation,sonnleitner2017will}).
Knowing the light field seen by the atom at the position $\bm r$ with velocity $\bm v = \bm p/m$ enables to solve the optical Bloch equations (density matrix $\hat \sigma (t)$ evolution) to determine the atomic internal state.
The force is then derived from $\bm F = - tr[\hat \sigma(t )\bm \nabla \hat H] =\displaystyle   \langle \frac{\partial \bm d.\bm E}{\partial \bm r} \rangle $. The usual optical Bloch equations  where $\sigma_{i j } (t)$ stands for $ \sigma_{i j } (t; r_0, v_0, t_0)$ read as

	\begin{equation}
	\begin{pmatrix}
	\frac{\partial   {\sigma}_{11} }{\partial t}	 & 	\frac{\partial  {\sigma}_{12} }{\partial t} \\ 
	\frac{\partial  {\sigma}_{21} }{\partial t} & 	\frac{\partial  {\sigma}_{22} }{\partial t}
	\end{pmatrix} (t) = - \frac{ 1}{2 i} \sum_{\rm L}
	\begin{pmatrix}
	\Upomega_{\rm L}^*(r(t),t)   {\sigma}_{21}(t) -  \Upomega_{\rm L}(r(t),t) {\sigma}_{12 } (t)   &  	\Upomega_{\rm L}^*(r(t),t)  ({\sigma}_{22}(t) -  {\sigma}_{11} (t)) \\ 
	\Upomega_{\rm L}(r(t),t)(   {\sigma}_{11}(t) -  {\sigma}_{22} (t)  ) & 
	\Upomega_{\rm L}(r(t),t)  {\sigma}_{12} (t)-  \Upomega_{\rm L}^* (r(t),t) {\sigma}_{21 } (t)
	\end{pmatrix} 
	\label{OBE}
	\end{equation}

where  $\displaystyle \Upomega_{\rm L}(r,t) = 
\Upomega_{\rm L} e^{ i ( {\bm k}_{\rm L}. {\bm r}  - \omega_{\rm L} t  - \Phi_{\rm L} )}$.
The rapidly oscillating terms  can be removed by introducing slowly varying quantities as
$\displaystyle  \sigma_{i j}^I (t) = e^{-i (E_j-E_i) t/\hbar} \sigma_{i j}(t)$. 


The absence of Doppler shift in the expression of $\Upomega_{\rm L}(r,t)$ may be surprising, especially when  compared to Eq. (\ref{eq_density})  (using $p'=p=p(t)$, $r=r(t)$ and  $\hbar k_{\rm L}$ put to $0$). 
The explanation is the following: we use $\bm r (t)= \bm r (t; \bm r_0, \bm v_0, t_0)$ so the Lagrangian description where individual particles are followed through time, whereas, when dealing with the Wigner $W(\bm r, \bm v, t)$ or PSD $\rho(\bm r, \bm v, t)$ picture, we use in the Eulerian description. 
The connection between Lagrangian and Eulerian coordinates explains why the Doppler effect is correctly taken in both Eq.(\ref{OBE}) with
$\displaystyle  \Upomega_{\rm L}(r(t),t) = 
\Upomega_{\rm L} e^{ i ( {\bm k}_{\rm L}. {\bm r(t)}  - \omega_{\rm L} t  - \Phi_{\rm L}(t) )} $,
and in 
Eq. (\ref{eq_density})
with	
$\displaystyle 
\Omega_{\rm L}(r,p,t)  = \Omega_{\rm L} e^{i (k_L r   +  k_{\rm L} p t /m  -  \delta_{\rm L}^0 t   - \Phi_{\rm L}(t))	}$. In any case, the instantaneous laser phase seen by the atoms is correct, including the Doppler effect because $\displaystyle  \frac{d r(t)}{d t} = p(t)  /m$.

Similarly,  in the Eulerian description the force  is thus given by $\displaystyle  {\rm Tr}[\hat \sigma(t )\bm \nabla \hat H]$, or $\displaystyle  {\rm Tr}[\hat \rho(t )^I \bm \nabla \hat V^I]$
using  the cyclic invariant of the  trace. We have
$\displaystyle  V^I(r,p,t) = - \sum_{\rm L} \frac{\hbar}{2} \Omega_{\rm L}(r,p,t)$ so
$\bm \nabla V^I(r,p,t) = \displaystyle 	 - i\sum_{\rm L} \frac{\hbar \bm k_{\rm L} }{2} \Omega_{\rm L}(r,p,t)$.

So in conclusion and back to our Lagrangian description we have:
\begin{equation}
\bm F (\bm r(t),\bm v(t), t) =\Im \left[ {\sigma}_{2 1 }(t)   	\sum_{\rm L} \hbar \bm k_{\rm L} \Omega_{\rm L}^*(\bm r(t),t)\right]
\label{force} 
\end{equation}
As we chose plane waves  (or $ \bm \nabla  E_{\rm L}=0$), there is no direct dipolar force. Also, because of the interplay between the Bloch equations (Eq.\ref{OBE}) and the force (Eq.\ref{force}), the
atomic velocity $\bm v(t )$ and position $\bm r(t )$ should be updated in a short time
interval (typically ps), and the calculation of the Bloch equation evolution iterated on a similar 
time scale 	
\cite{2016PhRvA..93f3410H}.

\subsection{Phase space evolution equation}
\label{Force_Wigner}

Here, we would like to justify the equations we just derived assuming a separation of the external and internal degrees of freedom.
However, we know that without spontaneous emission, this is valid 
only if the ratio of resonant photon momentum to atomic momentum dispersion is small $\hbar k/\Delta p \ll 1$. In such a case, 
the rapid processes acting on  the internal degrees of
freedom can be separated from the slow processes associated
with translational motion.
 The dynamics of the atomic ensemble is thus
determined by the slow change of the distribution function
in translational degrees of freedom $w(r, p)= W_{11}+W_{22}$ and 
the expansion in $\hbar k$, that we will derive here for completeness, is justified  \cite{dalibard1985atomic}.

One analogue of the classical phase space distribution
$\rho$ is
the total distribution function
in translational degrees of freedom, $w(r, p,t)$ as plotted in \cite[Fig. 2(b)]{main_art}.
Equations (\ref{Wigner_eq_brute_1}-\ref{Wigner_eq_brute_4}) (written for simplicity without the external potentials), become:
{\footnotesize
	\begin{eqnarray}
	\left[ \frac{\partial   }{\partial t}	 
	+\frac{ p}{m} \frac{\partial }{\partial  r}   
	\right]W_{11}(r,p,t) & =&   
	-
	\frac{1}{2 i} \sum_{\rm L} ( \Upomega_L^* (r,t)
	W_{21} (r,p+\hbar k_{\rm L}/2,t) - \Upomega_L (r,t) W_{12} (r,p+\hbar k_{\rm L}/2,t)
	) \label{eq_Wigner1}
	\\
	\left[	\frac{\partial   }{\partial t}+\frac{ p}{m}  \frac{\partial }{\partial  r} 
	-
	\frac{E_1 - E_2}{i\hbar}	
	\right] W_{12}(r,p,t) & = &
	-
	\frac{1}{2 i} \sum_{\rm L} \Upomega_L^* (r,t) (
	W_{2 2} (r,p+\hbar k_{\rm L}/2,t) - W_{11} (r,p-\hbar k_{\rm L}/2,t)
	) \label{eq_Wigner2}
	\\
	\left[	\frac{\partial   }{\partial t}  +\frac{ p}{m}  \frac{\partial }{\partial  r} 
	+
	\frac{E_1 - E_2}{i\hbar}
	\right] W_{21}(r,p,t) & = &
	-
	\frac{1}{2 i} \sum_{\rm L} \Upomega_L (r,t) (
	W_{11} (r,p-\hbar k_{\rm L}/2,t) - W_{22} (r,p+\hbar k_{\rm L}/2,t)
	) \label{eq_Wigner3}
	\\	
	\left[	\frac{\partial  }{\partial t}
	+\frac{ p}{m}  \frac{\partial }{\partial  r}   
	\right]   W_{22} (r,p,t)  & = & 
	- \frac{1}{2 i} \sum_{\rm L} ( \Upomega_L (r,t)
	W_{12} (r,p-\hbar k_{\rm L}/2,t) - \Upomega_L^* (r,t) W_{21} (r,p-\hbar k_{\rm L}/2,t)
	) \label{eq_Wigner4}
	\end{eqnarray}
}
with  
$\hbar \Upomega_{\rm L}(r,t) = 
\bm d. {\bm E}_{\rm L} e^{ i ( {\bm k}_{\rm L}. {\bm r}  - \omega_{\rm L} t  - \Phi_{\rm L} )} $.

An frequently used method to derive a continuity equation
as \cite[Eq.(1)]{main_art} for $\rho = w$
is to
expand  the Wigner distribution equations
in a power series of
the photon momentum $\hbar k$
\cite{minogin1987laser,kazantsev1990mechanical,dalibard1985atomic,1986RvMP...58..699S,1991JOSAB...8.1341Y,2003JETP...96..383B,2013JETP..117..222P}.
In the presence of spontaneous emission, the second order  leads to the standard
Fokker-Planck equation 	\cite{minogin1987laser,kazantsev1990mechanical,dalibard1985atomic,1986RvMP...58..699S,1991JOSAB...8.1341Y,2003JETP...96..383B,2013JETP..117..222P}. 
The simplest formulation is restricted to the first order approximation, therefore
$\displaystyle W_{21} (r ,p \mp \hbar k_{\rm L}/2,t) \approx	 W_{21} (r,p,t) \mp \frac{\hbar k_{\rm L}}{2}  \frac{\partial    }{\partial p}
\tilde W_{21} (r,p,t)$.
To this first order in $\hbar k_{\rm L}$, the sum of
(\ref{eq_Wigner1}) and (\ref{eq_Wigner4}) is:
	\begin{equation}
	\left[	\frac{\partial   }{\partial t}  +\frac{ p}{m}  \frac{\partial }{\partial  r} 
	\right]
	w(r,p,t)  =   
	-  \sum_{\rm L}
	\Im \left[   \Omega_L^* (r,t) 
	\hbar k_{\rm L}  \frac{\partial    }{\partial p}
	W_{21} (r,p,t)
	\right]
	\label{wigner_first_order}
	\end{equation}

Since the recoil momentum $\hbar k$ is small, the variation of
atomic translational motion is slower than the atomic internal state change. The latter follows the varying
translational state 	$w(r,p,t) $  \cite{minogin1984dynamics}.
Fast relaxation of the internal atomic state means that, the functions
$W_{ij} (r,p,t)$ follow the distribution function
$w(r,p,t) $. 

At zero order in $\hbar k_{\rm L}$ we  have the simplest approximation
$W_{ij} (r,p,t) \approx W_{ij}^0 (r,p,t) w(r,p,t)$.
Eq.(\ref{wigner_first_order}) leads to
\begin{equation}
\left[	\frac{\partial   }{\partial t}  +\frac{ p}{m}  \frac{\partial }{\partial  r} 
\right]
w(r,p,t)  =  - \frac{\partial   [ F ( r  ,  p, t)  w(r,p,t)] }{\partial p}
\end{equation}
We recognize a continuity equation as \cite[Eq.(1)]{main_art} with the force given by
\begin{equation} F ( r, p, t) =\Im \left[ W_{2 1 }^0( r, p,t)   	\sum_{\rm L} \hbar  k_{\rm L} \Omega_{\rm L}^* ( r,t)\right]
\label{force_wigner}
\end{equation}
So in a classical picture, this expression of the force shall be used to calculate individual particles trajectories.

The evolution of the Wigner function is given by Eqs.(\ref{eq_Wigner1})-(\ref{eq_Wigner4}), with 	$W_{ij} (r,p,t) \approx W_{ij}^0 (r,p,t) w(r,p,t)$, to obtain
\begin{widetext}
	\begin{eqnarray}	 
	\frac{\partial W_{11}^0(r+p t /m,p,t)  }{\partial t}	 & =&   
	- \sum_{\rm L}
	\Im \left[   	\Omega_{\rm L}^*(r+p t /m,t) 
	W_{21}^0 (r+p t /m,p,t)
	\right] 
	\\
	\frac{\partial W_{21}^0(r+p t /m,p,t)  }{\partial t}    & = &
	\frac{1}{2 i} \sum_{\rm L} 	\Omega_{\rm L}(r+p t /m,t)  (W_{22}^0(r+p t /m,p,t) -
	W_{11}^0 (r+p t /m,p,t)
	) \\	
	\frac{\partial 	W_{22}^0 (r+p t /m,p,t)  }{\partial t}
	& = & 
	\sum_{\rm L}
	\Im \left[   	\Omega_{\rm L}^*(r+p t /m,t)  W_{21}^0 (r+p t /m,p,t)
	\right]
	\end{eqnarray}
\end{widetext}
\onecolumngrid
where we have used
$\displaystyle	\left[ \frac{\partial   }{\partial t}	 
+\frac{ p}{m} \frac{\partial }{\partial  r}   
\right]W_{11}^0(r+p t /m,p,t) = 		\frac{\partial W_{11}^0(r+p t /m,p,t)  }{\partial t}$.	

We partially recognize the optical Bloch equations  (Eqs.\ref{OBE}), with $\sigma_{i j}(t) = 	W_{ij}^0 (r_0+p_0 t /m,p_0,t)$ \cite{dalibard1985atomic}. 
This is the usual   first order in time 
connection between  Lagrangian and Eulerian specification:
$\bm r (t)= \bm r (t; \bm r_0, \bm v_0, t_0) \approx r_0+v_0 t $,  $p(t) \approx p_0$.
So  to first order $\sigma_{i j}(t) \approx 	W_{ij}^0 (r( t),p(t),t)$ and
the force given by Eq.(\ref{force_wigner})  is exactly the same force as Eq.(\ref{force}). 

An alternative way to derive these expressions consists in using the interaction picture. A similar method using $w^I(r, p,t) = W_{11}^I+W_{22}^I$  
$W_{ij}^I (r,p,t) \approx {W_{ij}^I}^0 (r,p,t) w(r,p,t)$
from
Eqs.(\ref{int_wigner_a})-(\ref{int_wigner_c}) leads, to first order in $\hbar k_{\rm L}$ to:
\begin{widetext}
	\begin{eqnarray}	 
	\frac{\partial {W_{11}^I}^0  }{\partial t} (r,p,t)	 & =&   
	- \sum_{\rm L}
	\Im \left[   	\Omega_{\rm L}^*(r,p,t) 
	{W_{21}^I}^0 (r,p,t)
	\right] 
	\\
	\frac{\partial {W_{21}^I}^0  }{\partial t} (r,p,t)   & = &
	\frac{1}{2 i} \sum_{\rm L} 	\Omega_{\rm L}(r,p,t)  ({W_{22}^I}^0(r,p,t) -
	{W_{11}^I}^0 (r,p,t)
	) \\	
	\frac{\partial 	{W_{22}^I}^0   }{\partial t}(r,p,t)
	& = & 
	\sum_{\rm L}
	\Im \left[   	\Omega_{\rm L}^*(r,p,t)  {W_{21}^I}^0 (r,p,t)
	\right]
	\end{eqnarray}
\end{widetext}
\onecolumngrid

which are the usual Bloch equations in the particle frame. The Doppler effect is here explicitly included. Indeed, the continuity equation reads as
\begin{equation*}
\frac{\partial w^I  }{\partial t}	 
(r,p,t)  = -
\left[
- \frac{ t}{m} \frac{\partial    }{\partial r} 	 +  \frac{\partial    }{\partial p}\right] \left( F^I ( r, p, t) w^I(r,p,t) \right)  
\end{equation*}
for the force 
$F(\bm r + \bm p t /m,\bm p,t) = F^I ( r, p, t) =	\sum_{\rm L} \hbar k_{\rm L}   \Omega_L^* (r,p,t)
W_{21}^I (r,p,t)^0$.

This is indeed the classical continuity equation \cite[Eq.(1)]{main_art}. In the interaction picture 
$\rho(\bm r,\bm p,t) = \rho^I(\bm r-\bm p t /m,\bm p,t)$ leads to
\begin{equation}
\frac{\partial  \rho^I}{\partial t} (\bm r,\bm p,t)  + \left[
- \frac{ t}{m} \frac{\partial    }{\partial \bm r} 	 +  \frac{\partial    }{\partial \bm p}\right]   (\rho^I F^I)(\bm r,\bm p,t) = 0 \label{rho_rep_int}
\end{equation}
where 	$F(\bm r,\bm p,t) = F^I(\bm r-\bm p t /m,\bm p,t)$.

	\section{Defining quantitatively the PSD}
	We explicit the different quantities related to the generic term Phase Space Density (PSD) and Position Momentum Distribution (PMD) that are used in the core of the article:
		\begin{itemize}
\item	The PMD  are
 functions of position ($r$) and momentum ($p$).
\item  The PSD  are single values that are used to characterized how much the system is cold and dense.
 	\end{itemize}
 
 The PSD quantities  can be put into two main categories:
	\begin{enumerate}
		\item Position-momentum based PSD: will simply be the maximum of the PMD functions (such as the Wigner or Husimi distributions).
		\item Entropy based PSD: will simply be the value $D=e^{-S}$ for a given entropy $S$. The entropies are defined using  the density matrix $\hat \rho$. They  are of two types:
			\begin{itemize}
			\item Informational (or population-based, or diagonal) PSD: values linked to populations  $p_i = \langle i|  \hat \rho |i\rangle $ of specific states $|i\rangle$  (usually a complete basis set) chosen for their physical interest. 
		\item Eigenvalues (or spectral) PSD: values relying on eigenvalues $\lambda_i$ of the density matrix $\hat \rho$. 
			\end{itemize}
	\end{enumerate}
	
The PSD can include or not the internal states:
	\begin{itemize}
 \item For the full system, the PSD
is calculated from the whole density matrix of the full particle system $AB$ ($\hat \rho = \hat \rho_{AB}$) where $A$ and $B$ denote the subspaces related to the external and internal degrees of freedom respectively. Note that a quantification of the optical field would require a dedicated subspace $C$ and would lead to $\hat \rho = \hat \rho_{ABC}$. 

\item For the sole position-momentum, 
 we are only interested in the degrees of external freedom, i.e. coordinates $r,p$ regardless the internal degrees of freedom. Thus, the total density matrix is replaced by the partial density matrix obtained by tracing out the $B$ part: $\hat \rho_A  =Tr_B \hat \rho$. For instance, with a 2 level particle and a $|p,g/e\rangle $ basis, 	$ \langle p| {\hat \rho_A (t)}  |p' \rangle  =  \langle p,g| \hat \rho(t) |p',g \rangle +  \langle p,e| \hat \rho(t) |p',e \rangle $.
\end{itemize}
	
		\subsection{Position Momentum Distribution}

 The "usual" Wigner function  $W= W_{gg} + W_{ee}$, as plotted in Fig. 2 c) is given by  Eq. (\ref{def_Wigner}) with:
	\begin{eqnarray}
		W_{gg}(\bm r,\bm  p,t) &= & \frac{1}{h^3}
		\int \langle  \bm  p - \bm  p'/2,g | \hat \rho  | \bm  p + \bm p'/2,g \rangle e^{-i \bm r.\bm  p'/\hbar} d\bm  p'  
		\label{Wig}
	\end{eqnarray}
	and an equivalent expression for the excited state $W_{ee} $.

	A "smooth" version is obtained by averaging Eq. (\ref{Wig}) over an equivalent cell area of $2\pi \sigma_r \sigma_p $  weighted by a Gaussian function, which corresponds to the so called  Weierstrass transform (in 1D):
	$$
	W_G {\sigma_r,\sigma_p} (r,p) = \int dr' dp' W(r,p) G_{\sigma_r,\sigma_p}(r,r';p,p')
	$$
	where 
	$G_{\sigma_r,\sigma_p}(r,r';p,p') = \frac{2}{h} e^{\left( -\frac{(r-r')^2}{2 \sigma_r^2} - \frac{(p-p')^2}{2\sigma_p^2} \right)} $.
	$	W_{\sigma_r,\sigma_p} (r,p)$ represents a probability
	resulting from simultaneous measurement
	of position and momentum that is
	performed with a device whose uncertainties are $\sigma_r$ and $\sigma_p$  of is also used in this work \cite{1983SvPhU..26..311T,2012EPJST.203....3O,2013arXiv1303.3682M,2013arXiv1304.1034R,2016arXiv160303962M}.
	The Q-Husimi distribution is a special case with a minimal equivalent cell area of $h/2$ occurring when $ \sigma_r \sigma_p = \hbar/2$.
	This is the optimal distribution obtained for joint position and momentum  measurement \cite{2014JMP....55a2102R}. The	Husimi function is defined and positive and is equal to the average of the density operator over a	coherent state  $|\alpha(r,p) = \frac{r}{\sigma_r} + i  \frac{p}{\sigma_p} \rangle$. So,
$
	Q(r,p,t) = \frac{1}{\pi} \langle \alpha | \hat \rho_A | \alpha \rangle
	\label{Husimi}
$	
	is  the probability	distribution of the outcome of a heterodyne measurement performed on the state $|\alpha\rangle$ \cite{de2017wehrl}.
 $Q(r,p,t) = Q_{gg}(r,p,t)  + Q_{gg} (r,p,t) = \frac{1}{\pi} ( \langle \alpha,g |  \hat \rho | \alpha,g \rangle +  \langle \alpha,e |  \hat \rho | \alpha,e \rangle )$ is the function plotted in Fig. 2 d). Its maximum is plotted in Fig. 3.

		\subsection{Informational phase space density}
	
		Several states $|i\rangle$ can be used to define an informational PSD, such as energy states $|E_i\rangle $, momentum states $|p\rangle$ or also coherent states $|\alpha(r,p)\rangle$. For instance, if only the external degrees of freedom (subspace $A$) is of interest, the energy eigenstates are $E_p={\bm p}^2/2m$ for free particules,  $E_n = \hbar \omega (n+1/2) $ for 1D harmonically trapped particles. If, on the other hand, the full system $AB$ is considered, the internal energy must be added.

	 Several definitions of PSD are  possible
	 	depending of the choice  of the function of the parameters  $f(p_i)$ (see discussion below). An important one is the (Gibbs-)Shanon entropy $S_{\rm{Sh}} = -\,\sum_i p_i \ln \,p_i$. So, for the full space $AB$,
	\begin{equation}
	S_{\rm{Sh}} = - \big[ \sum_p \langle p,g |  \hat \rho |p,g\rangle \ln (\langle p,g |  \hat \rho |p,g\rangle) + \sum_p \langle p,e |  \hat \rho |p,e\rangle \ln (\langle p,e |  \hat \rho |p,e\rangle) \big]
	\end{equation}	
		  
while for the external degrees of freedom only,
 \begin{equation}
 S_{\rm{Sh}}^{\{A\}} = - \sum_p (\langle p,g| \hat \rho(t) |p,g \rangle +  \langle p,e| \hat \rho(t) |p,e \rangle) \ln (\langle p,g| \hat \rho(t) |p,g \rangle +  \langle p,e| \hat \rho(t) |p,e \rangle).
 \end{equation}
Finally, considering a specific internal state only, e.g. the ground state, it can be also defined 
 \begin{equation}
 S_{\rm{Sh}}^{(g)} = - \sum_p \langle p,g |  \hat \rho |p,g\rangle \ln (\langle p,g |  \hat \rho |p,g\rangle).
 \end{equation}

		   	  	\subsection{Spectral phase space density}
		   	  	
		   	  		The spectral PSD can be seen as a special case of population entropy when the state $|i\rangle$ are the eigenstates of the density matrix, i.e. $p_i=\lambda_i$. This gives rise to another definition of PSD known as Von Neumann entropy $S_{\rm VN}=- \sum_i \lambda_i \ln(\lambda_i)$. Such a definition has the advantage of being independent of the basis choice and is unambiguously defined from the density matrix as $S_{\rm VN}=	 - \mathrm{Tr}[ \hat{\rho} \ln(\hat{\rho}) ]$. 
The related PSD $ D_{\rm VN} =  e^{-S_{\rm VN}}$ was plotted for the full density matrix in Fig. 3 (a) and the partial density matrix in Fig. 3 (b).
The possible modification of $S_{\rm VN}^{\{A\}}$ is obviously linked to the mutual entropy $ S_{\rm VN}^{\{A\}} + S_{\rm VN}^{\{B\}} - S_{\rm VN}^{\{AB\}}$ defining the maximal cooling (work) that can be achieved  in quantum thermodynamics \cite{vinjanampathy2016quantum}.  The triangle inequality (Eq. 3 in the article) indicates that a subtly correlated system could even lead to an increase of $D_{\rm VN}^{\{A\}}$ by a factor $M^2$ \cite{bera2016universal}. However, under the canonical conditions where only one internal state is populated, the gain of $D_{\rm VN}^{\{A\}}$ is bounded to $M$ since $S_{\rm VN}^{\{AB\}}(0) = S_{\rm VN}^{\{A\}} (0) $ and $S_{\rm VN}^{\{AB\}}(t) = S_{\rm VN}^{\{AB\}}(0)$. This is consistent with the results shown in Fig. 3(b) where the gain on $D^{\{A\}}_{\rm{VN}}$ is greater than one but lower than $M=2$.
		   	  	  
		   	  	  \subsection{Other entropy definitions}
		   	  
Other functions $f$ of the parameters can be used to define the entropy.		   	  
For instance power function  leads to Tsallis entropy: $ S_q = \frac{1}{q-1} \left[ 1-\sum_i p_i^q \right]$. For $q \rightarrow 1$, it is reduced to the Shanon entropy  and for $q \rightarrow \infty$ to the maximal population of $\hat \rho$  (because $\displaystyle \lim_{q \rightarrow \infty} \|. \|_q =\|. \|_\infty $, that is $\displaystyle \lim_{q \rightarrow \infty} \bigg( \sum _i\left|p_{i}\right|^{q} \bigg)^{1/q}  = \max_i p_i$). 
		   	  	
		   	  Combining with  logarithmic function leads to the R\'enyi entropy $S_R^{(q)} = \frac{1}{1-q} \log \left[ \sum_i p_i^q \right]$. The case $q=0$ is the Hartley or max-entropy, $q \rightarrow 1$ is the Shannon entropy, $q=2$ is the Collision or simply called "R\'enyi" entropy and $q \rightarrow \infty$ the min-entropy.

		   	  	It is important to realize that for a given choice of $f$, a given PSD  will have a population  version $f(p_i)$ but also an eigenvalues  one (when $p_i=\lambda_i$). 
		   	  	Sometimes terminology is ambiguous and it is important to precise if we use a function of $p_i$ or $\lambda_i$. Fortunately, some definition are not   ambiguous, for instance
		   	  	the Von Neumann entropy is  always an eigenvalue one. The Von Neumann entropy is therefore always the Shannon entropy over the spectrum of $\hat \rho$.	Similarly the   so called (Tsallis-2) linear  entropy (because it approximates the Von Neumann  entropy when $\ln \hat \rho \approx \hat \rho -1$ \cite{wlodarz2003entropy})
		   	  	$S_L =   1-\sum_i \lambda_i^2  = 1- Tr({\hat \rho}^2)$  is  usually used  over the spectrum of $\hat \rho$ because it is  linked to the  measure of the purity of the quantum state (purity being  defined by $Tr(\hat{\rho}^2)$
		   	  	\cite{1997JChPh.106.1435B,TannorD.J._jp992544x}). 
		   	  	
		   	  		\subsection{Relation between PSD and PMD}

		   	  The  function $f$ can also be used to define a single value PSD from a PMD. 
		   	 For instance, we can define the so-called Wehrl entropy  
		   	  $ {\displaystyle S_{W} =-  \int Q(r,p)\ln Q(r,p)\,dr\,dp}$. This is a  continuous (or differential) entropy for $Q(r,p)$ seen as a probability density function. Wehrl's entropy is the classical limit $h \rightarrow  0$ of the Von Neumann quantum entropy \cite{beretta1984relation}. 
		   	  
		   	The linear  entropy could also be used  because, compared to 
		   	  other definitions of entropy, it has the privileged status 
		   	  to have a direct  Weyl-Wigner-Moyal transcription:
		   	  $S_R =  1- Tr({\hat \rho_A}^2) = 1- h \int W(r,p)^2 \,dr\,dp $  (so called Manfredi-Feix entropy) \cite{2000PhRvE..62.4665M,wlodarz2003entropy,2012PhRvA..86a2119S}.

   	  	\subsection{Relation and bounds  between PSD}
   	  	
   	  	\subsubsection{Informational versus spectral PSD}

   	  	A useful bound concern the fact that an informational entropy is always larger than the corresponding spectral entropy.
   	  	
   	The key argument is based on the Schur-Horn's theorem  (that indicates essentially that $p_i \leq \lambda_i$) and on the fact that,
   	in order to   to keep some basic properties of the entropies such as   increasing with disordered,  the functions $f$ are concave (so based on power or logarithmic functions).	
   	 Then Jensen's inequality for concave function proves the result
   	  \cite{gemmer2009quantum,2005PhRvA..71f3821B,2013LaPhy..23k5201S,frigg2011entropy,bengtsson2007geometry,vinjanampathy2016quantum,2018LMaPh.108...97D}. 
   	
   	 For instance $f(x) = - x \ln (x)$ leads to
  $	S_{\rm Sh} \geq S_{\rm VN}$
   	 (or  $	D_{\rm Sh} \leq D_{\rm VN}$).

   	 	\subsubsection{Invariance of full PSD}

  	 Invariance of full eigenvalues PSD are obvious using series of $f(x)$,	the  
  	 unitarity of the evolution operator $\hat U$ ($\hat \rho(t) = \hat U(t) \hat \rho(0) \hat{U}^\dagger$), and the cyclic
  	 invariant of the trace.
  	 The fact
  	 that all function of the eigenvalues $\lambda_i$ are conserved was the argument used in Ref. \cite{1992PhRvA..46.4051K}
  to mention that
  	 the	min entropy 
  	 $S_{\infty }=-\log \max _{i}\lambda_i =-\log \| \hat \rho \|_\infty$
  	 or the
  	 spectral radius $D_\infty =\| \hat \rho \|_\infty = \max_i(\lambda_i)$ of $\hat \rho$, that is the
  	 maximum
  	 occupation number of quantum eigenstates $\lambda_i $
  	 are  conserved under hamiltonian evolution (and so that the related PSD can no evolved). 
  	 
  	 	\subsubsection{Bound by the number of internal states $M$}

  	  The  evolution operator $\hat U$ can also be used to demonstrate some bounds \cite{1997JChPh.106.1435B} such as:
  	\begin{equation}
  \max \left[ {\hat \rho_A} (t) \right]  \leq M \max \left[ {\hat \rho_A} (0) \right] \label{max_rho}
  	\end{equation}
  	 
  	 That is demonstrated by considering $\max_p \left[ \langle p| {\hat \rho_A (t)}  |p \rangle \right] = \sum_{i=1}^M  \langle p_0,i| \hat \rho(t) |p_0,i \rangle $ 
  	 in addition to $  \langle p_0,i| \hat \rho(t) |p_0,i \rangle  =   \sum_{p,j}   U_{p_0 i,  p  j}  \rho_{pj,pj } (0)  U^*_{p j,  p_0 i} \leq  \max \left[ {\hat \rho_A (0)} \right] \sum_{p,j}   U_{p_0 i,  p  j}   U^*_{p j,  p_0 i} \leq  \max \left[ {\hat \rho_A (t)} \right] $ that arises from  the  unitarity of the evolution operator $\hat U$.

   	  	In a similar manner (using  $f(x)=x^n$ and   $\lim_{n \rightarrow \infty} \|. \|_n =\|. \|_\infty$ on  theorem 5 of \cite{2018LMaPh.108...97D}) it can be shown (see also \cite{de2017wehrl}) that the Husimi function $Q$ as well as the Wehrl entropy are bounded by the same factor $M$. More detail and other bounds can be find on Ref.
   	  	\cite{2013LaPhy..23k5201S,frigg2011entropy,gemmer2009quantum,2005PhRvA..71f3821B,bengtsson2007geometry,vinjanampathy2016quantum,2018LMaPh.108...97D}

    As an important final precaution, we mention that using pseudo phase space density definitions, as based on filtering of some specific states (such as for the ground state only $S_{\rm{Sh}}^{(g)} = - \sum_p \langle p,g |  \hat \rho |p,g\rangle \ln (\langle p,g |  \hat \rho |p,g\rangle)$), it is possible to find larger increase than a factor $M$. This is because such pseudo-PSD  are not based on a valuable density matrix.

\bibliographystyle{unsrt}

\end{document}